\documentclass{aa}
\usepackage{graphics}
\usepackage{psfig}
\usepackage{graphicx}
\usepackage{longtable}
\usepackage{latexsym}
\usepackage{amssymb}
\usepackage{lscape}

\begin{document}

   \title{ The far-ultraviolet emission of early-type galaxies  
    }

   \subtitle{}

   \author{J.-M. Deharveng
          \inst{}
          \and
          A. Boselli\inst{}
           \and
          J. Donas\inst{}
                  }

   \offprints{J.-M. Deharveng, \email{jean-michel.deharveng@astrsp-mrs.fr} }

   \institute{Laboratoire d'Astrophysique de Marseille, 
              Traverse du Siphon, Les Trois Lucs,
              BP 8, 13376 Marseille Cedex 12, France
                    }

   \date{Received  xxxx  2002 / Accepted xxxx 2002}

   \titlerunning{}
   \authorrunning{Deharveng et al.}

   \abstract{ We have assembled a UV-flux selected sample of 82 early-type 
       galaxies and collected additional information at other wavelengths.
       These data confirm a large spread of the $UV-V$ color in the range 
       2 to 5. The spread in $UV-V$ is 
       accompanied by a spread in $B-V$ that is mainly attributed 
       to the range of morphological types and luminosities. 
       A large fraction of the objects have red colors, $UV-V = 4 \pm 0.4$,
       corresponding to a weak UV-upturn as observed with IUE. 
       If the current interpretation for the UV emission from early-type
       galaxies is applicable to our sample, the PAGB (Post-Asymptotic 
       Giant Branch) tracks are the most common evolution path for the
       low-mass stars responsible for the UV emission. A small number of 
       very blue ($UV-V < 1.4$) objects have been found that 
       can be reasonably interpreted as harbouring some low level of
       star formation. In contrast with a previous sample 
       based on IUE observations, no correlation is found between the 
       $UV-V$ color and the Mg$_2$ spectral line index; 
       possible explanations are reviewed. The potential  
       of a more extended UV survey like GALEX is briefly presented.    
   \keywords{Stars: AGB and post-AGB -- Stars: horizontal-branch
               -- Galaxies: stellar content -- 
     Galaxies: elliptical and lenticular, cD -- Ultraviolet: galaxies}}

   \maketitle

%

\section{Introduction}

  The UV emission discovered in early-type galaxies
  as early as 1969 by the {\it Orbiting Astronomical Observatory-2}
  (Code et al. \cite{cod0}) is now currently interpreted 
  in terms of low-mass, helium-burning  stars in extreme horizontal branch and 
  subsequent stages of evolution.  O'Connell (\cite{oco}) has  
  extensively reviewed the built-up of that interpretation
  thanks to the combination of high quality UV data and new generations of 
  theoretical models for advanced
  stellar evolution (e.g. Greggio \& Renzini 
  \cite{gre1}, \cite{gre2} and references therein).
  The former include spectra with the Hopkins Ultraviolet 
  Telescope (HUT) (e.g. Ferguson \& Davidsen \cite{fer}, 
  Brown et al. \cite{bro0}) and high angular 
  resolution images with HST (e.g. Brown et al. \cite{bro2}, 
  \cite{bro3}). Notorious difficulties for converging on a well 
  accepted interpretation were the large variety of advanced stages of 
  stellar evolution and the sensitivity of UV production to small changes 
  of physical properties. 

      Because the most detailed observations are both time consuming and 
  intrinsically difficult, the current interpretation of the far-UV 
  radiation from early-type galaxies relies on a small number of
  objects. In addition to the bulge of M31, only 7 elliptical
  galaxies were spectroscopically observed with HUT; it is not yet 
  possible to resolve UV-bright stars down to the horizontal branch 
  beyond M31 (bulge) and its companions (Brown et al. \cite{bro3}). 
  Studying a large sample of 
  early-type galaxies would therefore require the use of cruder approaches,
  such as broad-band and integrated UV fluxes, but would 
  still be of interest.
  It would help to understand the generality of the conclusions 
  reached and to distinguish the possibility and frequency
  of low level of star formation in the population of
  early-type galaxies. Although this latter phenomenon is now excluded 
  as a general explanation for the UV emission, it may well be present 
  in a number of objects and have implications on galaxy evolution.

  A sample of 32 early-type galaxies was already studied by 
  Burstein et al. \cite{bur} (hereafter BBBFL) and, albeit observed 
  spectroscopically with IUE, was mostly discussed in terms of their
  ($1550-V$) color (see also Dorman et al. \cite{dor}). 
  With the availability of 
  several UV imaging surveys performed in the IUE-era
  (Brosch \cite{bros2}, O'Connell \cite{oco}), it is now possible
  to study a larger sample of early-type galaxies 
  in the far-UV.  Such a sample would have the advantage to 
  be essentially UV-flux selected and to potentially reveal UV emission 
  from unexpected early-type objects. This is a significant difference
  with the BBBFL sample made of objects with substantial
  record in the refereed literature and selected
  for one-by-one spectroscopic investigation.  
  An additional motivation of our approach is to prepare ourselves to 
  the extended UV survey of GALEX (Martin et al. \cite{mar}) and what  
  should be learnt of the early-type galaxies.

   The paper is organized as follows. Section 2 describes how our sample 
   of UV selected early-type galaxies has been built and 
   is complemented by a wealth of data at other wavelengths.  
   The $UV-V$ color distribution and color-color diagram 
   are presented in sections 3 and 4. The analysis follows in section 5. 
   We first take advantage of the fact that the BBBFL 
   sample contains most of the objects that have been studied in details
   to emphasize a possible relationship between the UV color and the 
   categories of stars responsible for the UV radiation. 
   We then discuss the relative frequencies of these categories of
   stars in the population of early-type galaxies, the possible 
   cases of recent star formation, the role of global 
   properties such as the luminosity,  
   the relation with the Mg$_2$ spectral line index and 
   the UV light profiles in a few objects.

\section {The sample}

\subsection{Data origin}

The sample analysed in this work is composed of all the optically selected 
early-type galaxies (type $\leq$ S0a) belonging to the Zwicky catalogue 
(CGCG, Zwicky et al. \cite{zwi}) ($m_{pg}$ $\leq$ 
15.7) detected in the UV by the FOCA experiment during the observations
of the Coma, A1367 and Cancer clusters (Donas et al. \cite{don2}, \cite{don3},
 private communication). 
To these, we add all the early-type galaxies belonging to the Virgo Cluster 
Catalogue (VCC; Binggeli et al. \cite{bin}) ($m_{pg}$ $\leq$ 18) detected by 
SCAP, FOCA and FAUST in the direction of Virgo (Donas et al. \cite{don1}, 
 private communication;
Deharveng et al. \cite{deh}). The sample is thus composed primarily of cluster 
galaxies, even though some background or foreground objects are also included.
Galaxies whose UV detection is doubtful because of confusion with nearby objects, 
(such as VCC 311), unless specified, have been systematically excluded.
The sample, largely dominated by objects observed with FOCA (85\%), 
is complete to a UV magnitude of about 18 whereas only 7 objects come from 
the less deep images of SCAP and FAUST.
Two additional early-type galaxies identified by Brosch et al. (\cite{bros1}) in 
their detailed study of FAUST images in the direction of the Virgo cluster have 
not been included. In order to preserve homogeneity and UV-flux selection,
the sample was not extended with other sources of UV data 
(UIT archives, O'Connell et al. \cite{oco0}; Maoz et al. \cite{mao}; 
Rifatto et al. \cite{rif} and references therein).

The final combined sample comprises 82 early-type 
galaxies, including a few dwarf ellipticals and spheroidals.
The accuracy of the morphological classification is excellent for the 
Virgo galaxies (Binggeli et al. \cite{bin}).
Because of the higher distance, the morphology of galaxies belonging 
to the other surveyed regions suffers from an uncertainty 
of about 1.5 Hubble type bins.\\

\subsection {UV data and precision}

  Most of the 
  UV data are total integrated magnitudes obtained at 2000 \AA~ with
  the FOCA experiment described by Milliard et al. (\cite{mil}).
  The FOCA UV magnitudes from Donas et al. (\cite{don2}, \cite{don3}) 
  have been reprocessed adopting a new zero-point calibration and 
  a revised version of the data reduction pipeline 
  (Donas et al., private communication).
   A comparison of the FOCA magnitudes with IUE data (stars and galaxies)
  has revealed large fluctuations from object to object, with the FOCA 
  fluxes being on average 0.3 mag brighter. Because of this dispersion 
  and various possible explanations on a case by case basis,
 we decided to stay on the FOCA calibration in order
 to be consistent with previous works.
  The comparison with IUE will again be addressed in the specific 
  context of the colors of the galaxies in section 3.
   The  UV magnitudes at 1650 \AA~ of the additional galaxies (6) from FAUST 
have been transformed to 2000 \AA ~using the relation 
$UV$(2000 \AA) = $UV$(1650 \AA) $+$ 0.2. This relation is intended to account for the
average spectral trend of ellipticals between 1650 \AA~ and 2000 \AA~ as well as 
the comparison of FAUST magnitudes with other UV measurements (Deharveng et al. 
 \cite{deh}). 
 The estimated error on the (FOCA) UV magnitude due to the flux extraction procedure
 and to the linearisation of the photographic plates is 
 0.3 mag in general, but it ranges from 0.2 mag for bright galaxies to 0.5 
 mag for weak sources. 
 This, combined with the previously discussed uncertainty on 
 the zero point, gives errors on the UV magnitudes of $\sim$ 0.5 mag.
 This uncertainty should be reminded when discussing color trends 
 in our sample; it is extremely large in comparison with the current range 
 of optical colors (as $B-V$) but should be seen in the context 
 of the much larger range of variation of the UV color. 
 A comparison of 4 galaxies measured with both FOCA and FAUST (for homogeneity
 only the FOCA data have been retained in the sample) shows the FOCA fluxes 
 0.55 mag fainter than FAUST fluxes on average. This number suggests a  
 possible systematic effect but remains consistent with our
 evaluation of the uncertainty of UV magnitudes.  

\subsection{Complementary data}

Optical data, available for 63 objects in the V, 72 in the B and 51 in the U band
are from Gavazzi \& Boselli (\cite{gav1}) and Boselli et al. (private communication).
NIR data, from Nicmos3 observations, are taken from Boselli et al. (\cite{bos1}) 
and Gavazzi et al. (\cite{gav5}, \cite{gav7}) (74 galaxies). 
From these data we derive total (extrapolated to infinity)
magnitudes $H_T$ determined as described in Gavazzi et al. (\cite{gav6})
with typical uncertainties of $\sim$ 10 \%.  For a few objects we
derive the H luminosity
from K band measurements assuming an average $H-K$ color of
0.25 mag (independent of type; see Gavazzi et al. \cite{gav6}) when the true $H-K$
color is not available.
The estimated error on the optical and near-IR magnitudes is 0.1 mag. 

The multifrequency data used in this work are listed in Table 1, 
arranged as follow:

\begin{itemize}

\item{Column 1: VCC designation, from Binggeli et al. (\cite{bin}) for Virgo galaxies,
or CGCG (Zwicky et al. \cite{zwi}) for A1367, Coma and Cancer cluster galaxies.}
\item{Column 2: UGC name.}
\item{Column 3: NGC/IC name.}
\item{Column 4: morphological type as given in the VCC for Virgo galaxies or 
  in Gavazzi \& Boselli (\cite{gav1}) for the other objects.}
\item{Column 5: photographic magnitude from the VCC for Virgo galaxies, 
                from the CGCG for the other objects.}
\item{Columns 6 and 7: major and minor optical diameters. For VCC galaxies the
diameters are measured on the du Pont plates at the faintest detectable isophote. 
For CGCG galaxies these are the major and minor optical diameters 
($a_{25}$, $b_{25}$) (in arcmin) derived as explained in Gavazzi \& Boselli (\cite{gav1})}
\item{Column 8: distance, in Mpc. Distances to the various 
substructures of Virgo are as given in Gavazzi et al. (\cite{gav3}). 
  A distance of 91.3 and 96 Mpc is assumed 
for galaxies in the A1367 and Coma clusters respectively. 
For Cancer cluster galaxies and for background and foreground objects the distance is determined
from the redshift assuming $H_0$= 75 $\rm km~s^{-1}Mpc^{-1}$.}
\item{Column 9: cluster membership as defined in Gavazzi et al. (\cite{gav3}) for Virgo 
   and in Gavazzi et al. (\cite{gav4}) for A1367 and Coma. P is for pairs, G for groups, 
 BkgV for galaxies in the background of Virgo,
 ForC for objects in the foreground of Coma.}
\item{Columns 10 to 15: K, H, J, V, B and U magnitudes determined as in 
  Gavazzi \& Boselli (\cite{gav1}), 
corrected for galactic extinction according to Burstein \& Heiles (\cite{burstein}). 
 S0a galaxies are corrected
for internal extinction as in Gavazzi \& Boselli (\cite{gav1}).}
\item{Column 16 and 17: $UV$ (2000 \AA) magnitude corrected for galactic extinction according 
to Burstein \& Heiles (\cite{burstein}) assuming $A(UV)= 2.1 \times A(B)$ 
 (all the targets being high galactic 
latitude objects, $A(UV) \leq$ 0.3 mag), and reference.}
\item{Column 18: Mg$_2$ data for the nuclear regions from Golev \& Prugniel 
 (\cite{gol}) and Jorgensen (\cite{jor}). The index is defined as in Worthey 
 (\cite{wor}).}
\item{Column 19: logarithm of the H band luminosity, in solar units, determined from 
the relation log$L_H$ = 11.36 + 2log$D$ $-$ 0.4$H_T$, where $H_T$ is the total extrapolated 
 $H$ magnitude
and $D$ the distance (in Mpc).}
\item{Column 20: the $C_{31}$ index, defined as the ratio of the radii containing 75\% to 25\%
of the total H-band light of the galaxy.}
\item{Column 21: comments to individual objects}
\end{itemize}

\addtocounter{table}{0}
\onecolumn
\clearpage
\scriptsize
\begin{landscape}
\begin{longtable}{rcccccccccccccccccccc}
\caption{The sample galaxies: Virgo}\\
\hline
\hline
\noalign{\bigskip}
 VCC  &  UGC &  NGC/IC &  type &  m$_{pg}$ &  a &  b &  Dist &  Cluster &
 Kmag &  Hmag &  Jmag &Vmag &   Bmag & Umag& UVmag & Ref &  Mg$_2$ &  log L$_H$ &  C$_{31}$&Note\\
\noalign{\smallskip}
(1) & (2) & (3) & (4) & (5) & (6) & (7) & (8) & (9) & (10) & (11) & (12) & (13) & (14) & (15) & (16) & (17) & (18)
&(19)&(20)&(21)\\
\hline
\noalign{\smallskip}
       49 & 7203&4168&  E&12.21& 1.76& 1.40&   32&    M & 8.32& 8.84& 9.62&11.72&12.63&13.11&12.50& 2& 0.246&   10.89&    3.71&*\\
      288 &    -&   -& dE&17.70& 0.52& 0.31&   17&    N &  -  &  -  &   - &  -  &  -  &  -  &16.82& 8&   -  &     -  &     -  &\\
      355 & 7365&4262& S0&12.41& 1.87& 1.63&   17&    A & 8.37& 8.59& 9.33&11.62&12.56&13.10&14.93& 8& 0.294&   10.38&    5.72&\\
      389 &    -& 781&dS0&14.21& 1.41& 0.91&   17&    A &  -  &  -  &  -  &  -  &  -  &  -  &16.85& 8&   -  &     -  &     -  &\\
      608 &    -&4322& dE&14.94& 1.25& 0.62&   17&    A &11.80&11.70&12.62&14.52&15.24&15.46&16.89& 8&   -  &    9.07&    2.66&\\
      616 &    -&4325&  E&14.40& 1.55& 0.97&102.8& BkgV &  -  &  -  &  -  &13.47&14.40&14.85&14.06& 1&   -  &     -  &     -  &\\
      715 &    -&3274& S0&14.80& 0.93& 0.46& 90.9& BkgV &  -  &  -  &  -  &  -  &  -  &  -  &17.27& 9&   -  &     -  &     -  &\\
      731 & 7488&4365&  E&10.51& 8.73& 6.18&   23&    B & 6.50& 6.78& 7.48& 9.66&10.64&11.25&13.75& 1& 0.312&   11.42&    6.00&*\\
      759 & 7493&4371& S0&11.80& 5.10& 2.48&   17&    A & 7.77& 8.05& 8.76&10.87&11.85&12.41&15.61& 6&   -  &   10.62&    3.93&\\
      763 & 7494&4374&  E&10.26&10.07&10.07&   17&    A & 6.43& 6.69& 7.43& 9.16&10.16&10.76&13.71& 6& 0.287&   11.16&    4.70&*\\
      781 & 7500&3303&dS0&14.72& 1.08& 0.50&   17&    A &12.11&  -  &  -  &14.34&15.03&  -  &16.92& 6&   -  &    8.94&    3.04&*\\
      828 & 7517&4387&  E&12.84& 1.84& 0.83&   17&    A & 9.04& 9.32&10.03&12.29&13.19&13.74&16.67& 6& 0.228&   10.08&    4.25&\\
      870 &    -&3331&dS0&15.52& 1.16& 0.43&185.7& BkgV &  -  &  -  &  -  &  -  &  -  &  -  &16.33& 6&   -  &     -  &     -  &\\
      881 & 7532&4406&  E&10.06&11.37& 7.51&   17&    A & 6.04& 6.27& 6.98& 8.95& 9.94&10.51&14.11& 6& 0.290&   11.38&    7.06&*\\
      914 &    -&   -& dE&19.00& 0.25& 0.25&   23&    B &  -  &  -  &  -  &  -  &  -  &20.02&17.29& 9&   -  &     -  &     -  &\\
      944 & 7542&4417& S0&12.08& 3.60& 1.00&   23&    B & 8.21& 8.42& 9.20&11.13&12.03&12.52&15.58& 9&   -  &   10.89&   12.52&*\\
      951 & 7550&3358&dE/dS0&14.35& 1.43& 0.94&17&    B &  -  &11.54&  -  &13.94&14.58&  -  &16.32& 6&   -  &    9.18&    3.25&\\
     1003 & 7568&4429&S0a&11.15& 8.12& 3.52&   17&    A & 6.54& 6.74& 7.43& 9.59&10.58&11.22&14.98& 6& 0.232&   11.01&    5.48&*\\
     1010 & 7569&4431&dS0&13.68& 1.58& 0.79&   17&    A &10.53&10.74&11.35&13.23&14.05&14.59&16.91& 6& 0.153&    9.57&    2.86&\\
     1030 & 7575&4435& S0&11.84& 2.92& 2.48&   17&    A & 7.68& 7.92& 8.66&10.94&11.82&12.33&15.31& 6& 0.189&   10.74&   10.06&*\\
     1111 &    -&   -& dE&17.70& 0.33& 0.20&   17&    A &  -  &  -  &  -  &  -  &  -  &  -  &15.73& 6&   -  &     -  &     -  &\\
     1125 & 7601&4452& S0&13.30& 2.92& 0.57&   17&    A & 9.04& 9.31&10.10&11.91&12.87&13.41&16.23& 6&   -  &   10.06&    2.67&\\
     1146 & 7610&4458&  E&12.93& 1.80& 1.52&   17&    A & 9.38& 9.69&10.32&12.32&13.18&13.64&16.74& 6& 0.208&   10.06&    7.99&\\
     1226 & 7629&4472&  E& 9.31&10.25& 8.11&   17&    S & 5.30& 5.59& 6.31& 8.54& 9.52&10.16&13.35& 9& 0.313&   11.66&    7.34&*\\
     1250 & 7637&4476& S0&12.91& 1.89& 0.94&   17&    A & 9.50& 9.82&10.47&12.41&13.23&13.55&15.33& 6& 0.141&    9.88&    3.59&\\
     1279 & 7645&4478&  E&12.15& 1.89& 1.43&   17&    A & 8.24& 8.52& 9.17&11.45&12.36&12.77&15.47& 6& 0.233&   10.35&    3.37&\\
     1297 &    -&   -&  E&14.33& 0.51& 0.45&   17&    A &10.07&10.34&11.13&13.44&14.42&15.03&17.41& 6& 0.290&    9.72&    3.52&\\
     1316 & 7654&4486&  E& 9.58&11.00&11.00&   17&    A & 5.92& 6.19& 7.01& 8.82& 9.82&10.37&12.70& 6& 0.270&   11.34&    4.20&*\\
     1327 & 7658&   -&  E&13.26& 1.10& 0.88&   17&    A & 9.07& 9.23& 9.75&11.49&12.13&12.27&14.49& 6&   -  &   10.14&    4.43&*\\
     1368 & 7665&4497&S0a&13.12& 2.01& 0.85&   17&    A & 9.60& 9.72&  -  &12.18&13.05&13.46&17.22& 6&   -  &    9.85&    2.81&\\
     1499 &    -&3492&  E&14.94& 0.64& 0.46&   17&    A &  -  &12.59&  -  &14.77&15.26&  -  &13.79& 1&   -  &    8.79&    2.74&*\\
     1535 & 7718&4526& S0&10.61& 7.00& 2.01&   17&    S & 6.37& 6.65& 7.47& 9.83&10.80&11.36&14.04& 1& 0.272&   11.19&   10.59&\\
     1809 & 7825&3631&S0a&14.17& 1.10& 0.67& 37.3& BkgV &  -  &  -  &  -  &  -  &  -  &  -  &10.70& 1&   -  &     -  &     -  &\\
\hline
\newpage
\caption{The sample galaxies: A1367, Cancer, Coma}\\
\hline
\noalign{\bigskip}
CGCG  &  UGC &  NGC/IC &  type &  m$_{pg}$ &  a &  b &  Dist &  Cluster &
 Kmag &  Hmag &  Jmag& Vmag &   Bmag & Umag& UVmag & Ref &  Mg$_2$ &  log L$_H$ &  C$_{31}$&Note\\
\noalign{\smallskip}
(1) & (2) & (3) & (4) & (5) & (6) & (7) & (8) & (9) & (10) & (11) & (12) & (13) & (14) & (15) & (16) & (17) & (18)
&(19)&(20)&(21)\\
\hline
\noalign{\smallskip}
    97125 &    -&   -&S0a&15.60& 0.84& 0.59& 91.3&A1367 &  -  &11.82&  -  &14.67&15.55&15.82&16.05& 3&   -  &   10.67&    5.76&*\\
    97127 & 6723&3862&  E&14.00& 1.62& 1.58& 91.3&A1367 & 9.75&10.00&10.72&12.90&13.86&14.43&16.32& 5& 0.282&   11.32&    5.88&*\\
    97134 & 6731&3867& S0&14.60& 1.31& 0.44& 91.3&A1367 &  -  &10.36&  -  &13.32&14.28&14.86&18.04& 3&   -  &   11.45&    4.02&\\
    98078 &    -&   -&  E&15.20& 0.40& 0.30& 91.2& P    &  -  &12.35&  -  &  -  &  -  &  -  &13.25& 1&   -  &   10.38&     -  &*\\
   119030 &    -&   -&  E&15.70& 0.66& 0.44& 31.2&Cancer&  -  &13.07&  -  &15.45&16.03&15.98&16.72& 7&   -  &    9.03&    3.65&*\\
   119053 &    -&   -&S0a&15.50& 0.63& 0.50& 66.4&Cancer&12.17&12.47&12.77&14.80&15.30&15.28&14.87& 7&   -  &   10.12&    8.67&*\\
   119065 & 4347&2563&  E&13.70& 2.60& 2.21& 66.4&Cancer&  -  & 9.46&  -  &12.13&13.04&13.63&16.01& 7& 0.313&   10.98&    9.44&\\
   119086 &    -&   -&  E&15.70& 0.53& 0.40& 89.2&Cancer&  -  &13.47&  -  &15.39&15.77&  -  &15.07& 7&   -  &    9.84&    2.86&\\
   127032 & 6663&3821& S0&13.80& 1.77& 1.51& 91.3&A1367 &  -  &10.56&  -  &12.66&13.32&13.59&15.03& 3&   -  &   11.11&    5.45&*\\
   127045 & 6725&   -&S0a&14.50& 1.50& 1.20& 91.3&A1367 &  -  &11.15&  -  &  -  &  -  &  -  &16.18& 5&   -  &   10.94&    4.35&*\\
   127048 &    -&   -&  E&15.00& 0.50& 0.50& 93.4& G    &  -  &11.20&  -  &  -  &  -  &  -  &16.49& 5&   -  &   10.91&    7.05&\\
   160014 &    -&   -&  E&15.70& 0.80& 0.54&   96& Coma &  -  &11.79&  -  &  -  &15.38&  -  &16.56& 5&   -  &   10.58&    3.29&\\
   160021 & 8057&4816& S0&14.80& 2.04& 1.41&   96& Coma & 9.89&10.06&10.87&12.77&13.74&14.18&16.89& 4& 0.304&   11.32&    8.21&\\
   160028 & 8065&4827& S0&14.10& 1.50& 1.16&   96& Coma &10.08&10.37&  -  &13.16&14.10&  -  &16.92& 5& 0.327&   11.27&    8.71&\\
   160038 & 8069&   -&S0a&14.80& 1.18& 0.53&   96& Coma &  -  &10.92&  -  &13.57&14.35&15.03&17.56& 5&   -  &   10.93&    5.84&\\
   160039 & 8070&4839&  E&13.60& 3.56& 1.55&   96& Coma & 9.60& 9.85&10.58&12.15&13.15&13.71&16.27& 5& 0.312&   11.53&   10.20&*\\
   160042 &    -&4840&  E&14.80& 0.95& 0.85&   96& Coma &10.56&10.84&11.58&13.78&14.73&15.32&17.45& 4& 0.323&   11.08&    8.73&\\
   160044 & 8072&4841&  E&13.50& 1.59& 1.55&   96& Coma & 9.74& 9.90&10.70&12.75&13.72&  -  &16.82& 4& 0.317&   11.45&    8.72&*\\
   160059 &    -&   -&  E&15.20& 1.31& 0.25&   96& Coma &  -  &11.59&  -  &  -  &15.16&  -  &17.85& 5&   -  &   10.74&    4.51&\\
   160063 &    -&4850& S0&15.30& 0.79& 0.61&   96& Coma &11.24&11.54&12.34&14.29&15.22&15.81&16.90& 4& 0.287&   10.75&    3.82&\\
   160068 & 8092&4853& S0&14.20& 1.00& 0.78&   96& Coma &10.45&10.91&11.61&13.61&14.30&14.55&15.42& 4& 0.164&   10.98&    3.77&*\\
   160077 &    -&3990&S0a&15.00& 1.22& 0.49&   96& Coma &  -  &10.61&  -  &  -  &14.37&  -  &17.48& 4&   -  &   11.05&    7.55&\\
   160079 &    -&   -&S0a&15.10& 0.98& 0.38&   96& Coma &10.87&11.18&11.76&  -  &14.56&  -  &17.97& 5& 0.251&   10.83&    5.48&\\
   160100 &    -&   -&  E&15.50& 0.72& 0.67&   96& Coma &11.51&11.84&12.55&14.73&15.65&  -  &17.87& 4& 0.279&   10.62&    4.89&\\
   160101 &    -&   -&S0a&15.20& 1.03& 0.35&   96& Coma &10.95&11.24&11.88&  -  &14.84&  -  &17.90& 4& 0.286&   10.79&    6.15&\\
   160103 & 8142&4926&  E&14.10& 1.31& 0.99&   96& Coma & 9.91&10.25&10.93&13.10&14.08&  -  &17.29& 4& 0.315&   11.28&    8.11&\\
   160104 &    -&   -&S0a&15.40& 0.67& 0.28&   96& Coma &12.26&12.55&13.12&  -  &15.26&  -  &16.49& 4&   -  &   10.28&    3.95&*\\
   160105 &    -&4927& S0&14.80& 1.07& 0.74&   96& Coma &10.35&10.69&11.50&13.72&14.73&  -  &17.34& 4& 0.348&   11.14&    8.42&\\
   160107 &    -&   -&S0a&14.90& 1.06& 0.32&   96& Coma &  -  &10.66&  -  &13.50&14.51&15.78&17.39& 5&   -  &   10.99&    3.66&\\
   160109 &    -&   -& S0&15.50& 0.67& 0.54&   96& Coma &11.29&11.48&12.36&  -  &15.45&  -  &17.42& 4&   -  &   10.75&    4.53&\\
   160118 & 8154&4931& S0&14.40& 1.76& 0.65&   96& Coma &10.22&10.45&11.15&13.35&14.22&  -  &16.95& 4&   -  &   11.22&    8.81&\\
   160120 & 8160&4934& S0&15.00& 1.33& 0.36&   96& Coma &  -  &11.58&  -  &14.36&15.11&15.55&17.40& 4&   -  &   10.76&    3.89&\\
   160122 &    -&   -& S0&15.60& 0.71& 0.53&   96& Coma &10.39&11.69&  -  &  -  &15.50&  -  &17.45& 4&   -  &   10.62&    4.05&\\
   160124 & 8167&4944& S0&13.30& 2.32& 0.84&   96& Coma &10.15&10.31&11.05&12.87&13.78&14.12&16.53& 4&   -  &   11.22&    5.76&\\
   160125 &    -&   -& S0&15.40& 0.89& 0.69&   96& Coma &  -  &11.52&  -  &  -  &15.46&  -  &17.75& 5&   -  &   10.73&    7.31&\\
   160129 & 8175&4952&  E&13.60& 1.74& 1.18& 78.2& ForC &  -  &10.08&  -  &12.94&13.79&  -  &16.68& 5& 0.290&   11.23&    8.46&\\
   160140 &    -&4971& S0&15.00& 0.99& 0.86& 85.3& ForC &10.80&10.96&  -  &14.01&14.94&  -  &17.32& 5& 0.284&   10.94&    9.25&\\
   160211 &    -&3947& S0&15.60& 0.58& 0.51&   96& Coma &11.86&12.03&12.79&14.91&15.77&  -  &17.38& 4& 0.282&   10.56&    4.14&\\
   160222 &    -&4867&  E&15.50& 0.64& 0.43&   96& Coma &11.29&11.68&12.33&14.53&15.45&15.92&16.73& 4& 0.294&   10.69&    3.92&\\
   160228 &    -&3973& S0&15.20& 0.89& 0.50&   96& Coma &11.02&11.34&12.17&14.30&15.28&15.81&18.12& 5& 0.318&   10.80&    7.40&\\
   160229 &    -&4873& S0&15.40& 0.57& 0.49&   96& Coma &11.48&11.62&12.53&14.44&15.41&15.94&18.13& 5& 0.287&   10.77&    5.11&\\
   160231 & 8103&4874&  E&13.70& 2.27& 1.93&   96& Coma & 8.91& 9.22&10.02&12.02&12.97&13.55&16.50& 4& 0.310&   11.81&    5.19&*\\
   160234 &    -&4876&  E&15.10& 0.61& 0.52&   96& Coma &11.30&11.57&12.37&14.52&15.47&16.03&18.36& 5& 0.260&   10.71&    4.11&\\
   160241 & 8110&4889&  E&13.00& 3.30& 2.23&   96& Coma & 8.29& 8.67& 9.41&11.48&12.44&13.06&15.58& 4& 0.342&   11.68&    4.35&*\\
   160248 &    -&4898&  E&14.70& 0.85& 0.62&   96& Coma &10.82&11.03&11.75&13.68&14.63&15.17&17.27& 5& 0.283&   10.89&    8.97&\\
   160249 & 8113&4895& S0&14.30& 2.00& 0.66&   96& Coma &10.27&10.53&11.31&13.19&14.12&14.64&16.74& 4& 0.301&   11.28&   12.39&*\\
   160256 &    -&4045&  E&15.10& 0.84& 0.64&   96& Coma &10.98&11.27&11.94&14.01&14.99&15.48&17.97& 5& 0.299&   10.82&    3.60&\\
   160258 &    -&4908&  E&14.90& 0.92& 0.68&   96& Coma &10.60&10.93&11.61&13.83&14.79&15.27&17.71& 5& 0.282&   11.05&    8.20&\\
   160259 & 8129&4051&  E&14.80& 1.49& 0.98&   96& Coma &10.36&10.78&11.51&13.46&14.40&14.96&17.10& 4& 0.331&   11.07&    4.48&\\

\noalign{\smallskip}
\hline              
\end{longtable} 
\end{landscape}
\footnotesize{ 
Comments to individual objects:

Virgo:

49: low-luminosity dwarf Seyfert nucleus (Ho et al. \cite{ho}) 

731: $H\alpha+[NII]$E.W.=2 \AA~ (Kennicutt \& Kent \cite{ken0})

763: Low Excitation Radio Galaxy (NED); $H\alpha+[NII]$E.W.=1 \AA~ 
(Trinchieri \& Di Serego Alighieri \cite{tri});
         also measured with FAUST (Deharveng et al. \cite{deh}) 

781: interacting with NGC 4388? (Corbin et al. \cite{cor})

881: M86; $H\alpha+[NII]$E.W.=13 \AA~ (Trinchieri \& Di Serego Alighieri \cite{tri});
      also measured with FAUST (Deharveng et al. \cite{deh}) 

944: classified `` SB0: sp'' in NED

1003: $H\alpha+[NII]$E.W.=5 \AA~ (Boselli \& Gavazzi \cite{bos2})

1030: interacting with NGC 4438; LINER

1226: M49; $H\alpha+[NII]$E.W.=1 \AA~ (Kennicutt \& Kent \cite{ken0});
               also measured with FAUST (Deharveng et al. \cite{deh})  

1316: M87; radio galaxy; $H\alpha+[NII]$E.W.=2 \AA~ (Boselli \& Gavazzi \cite{bos2});
             also measured with FAUST (Deharveng et al. \cite{deh}) 

1327: bright star superposed (Prugniel et al. \cite{pru}) 

1499: unresolved in UV from the nearby companion VCC 1491; given the extreme red color of 
 VCC 1491 and the blue color of VCC 1499
(Gavazzi et al. \cite{gav7}), the UV source should be identified to VCC 1499 (instead of 
VCC 1491 in Deharveng et al. \cite{deh}). VCC 1499 is as blue as a dwarf irregular and 
has a spectrum with the characteristics of a post-starburst galaxy (PSB) (Gavazzi et al.
 \cite{gav7}). 

Coma/A1367/Cancer:

97125: $H\alpha+[NII]$E.W.=26 \AA~ (Moss et al. \cite{mos}), 
        $H\alpha+[NII]$E.W.=21 \AA~ (Gavazzi et al. \cite{gav2})

97127: Low Excitation Radio Galaxy (NED)

98078: Mrk 758, $H\alpha+[NII]$E.W.=82 \AA~ (Gavazzi et al. \cite{gav2})                  

119030: classified ``spiral'' in NED

119053: $H\alpha+[NII]$E.W.=42 \AA~ (Kennicutt et al. \cite{ken})

127032: classified as (R)SAB(s)ab in NED

127045: $H\alpha+[NII]$E.W.=13 \AA~ (Moss et al. \cite{mos})

160039: radio galaxy (NED)

160044: binary system

160068: AGN (NED); Balmer absorption lines (Sparke et al. \cite{spa}); 
    PSB (Caldwell et al. \cite{cal})

160104: PSB (Caldwell et al. \cite{cal1})

160231: cD galaxy

160241: cD galaxy

160249: classified as ``SA0 pec sp'' in NED

References to the UV data:
1: FAUST data (Deharveng et al. \cite{deh});
2: SCAP data (Donas et al. \cite{don1});
3: Donas et al. \cite{don2} (reprocessed data, $\Delta$zero-point=$+$0.30 mag, see text);
4: Donas et al. \cite{don3} (reprocessed data, $\Delta$zero-point=$-$0.12 mag, see text);
5--9: Donas et al. private communication (5, Coma-A1367), (6, Virgo center),
  (7, Cancer), (8, Virgo M100), (9, Virgo M49)}
\normalsize
\twocolumn

\section{The $UV-V$ color distribution}

        The scatter of the early-type galaxies in the UV in comparison 
   with their behavior in the optical to near-infrared domain 
   was noted as
   early as the first {OAO-2} observations 
   (Code \& Welch \cite{cod1}).
   This scatter was confirmed by subsequent
   observations (e.g. Burstein et al. \cite{bur}),
   and it was shown as well that 
   the scatter among UV colors of galaxies 
   with similar types was decreasing from 
   early to late types (Code \& Welch \cite{cod2}, Smith \& Cornett 
  \cite{smi},  Deharveng et al. \cite{deh}). 
 
     The distribution of the $UV-V$ colors of 63 galaxies 
   of our sample are illustrated in Figure 1 
   and compared with that of the ($1550-V$) colors  
   of BBBFL. The distributions are similar in several aspects:
   a peak near a color of 4, a relatively sharp cut-off on the red side, 
   a more shallow decrease on the blue side (from color 4 to 2) 
   and a few ``outliers'' on the very blue side (precise definition and 
   implications to be seen later on). In both cases the bulk of the 
   objects have colors in the range 2 to 5 or so. 

      For five objects in common (NGC 4374, NGC 4406, NGC 4472, NGC 4486, NGC 4889),
  our $UV-V$ colors are found 1.4 mag redder on average than those of BBBFL. Three
  factors may account for this disturbingly large difference. First, 
  our ($2000-V$) colors
  are less influenced than the ($1550-V$) colors by the UV-upturn phenomenon. 
  Second, our measurements are integrated whereas the IUE data are 
  for a 10\arcsec $\times$
  20\arcsec~ region centered on the nucleus, and all objects (except M32) are known 
  to become redder in $UV-B$ at large radii (Ohl et al. \cite{ohl}). Last but not least,
  the UV flux is integrated in a smaller area than the V light   
  in the case of objects that are well resolved with FOCA; for 3 of the
  five objects in common, this factor would reduce the $UV-V$ color by about 0.7 mag.
  For this reason and since the peak of our histogram (Figure 1) 
  appears only slightly redder than the BBBFL distribution,
  we remain confident that a large discrepancy in $UV-V$ colors
  is not the rule.

   It is important to note that the increase in the number of objects 
   from the blue to the peak at a color of $\sim$ 4 is opposite to 
   the trend expected from selection effects with 
   UV-flux limited samples. These effects decrease the 
   volume for finding red galaxies, hence their proportion
   relative to blue galaxies. This very trend certainly plays a role 
   in the cutoff at color $>$ 4.  

\begin{figure}
  \resizebox{\hsize}{!}{\includegraphics{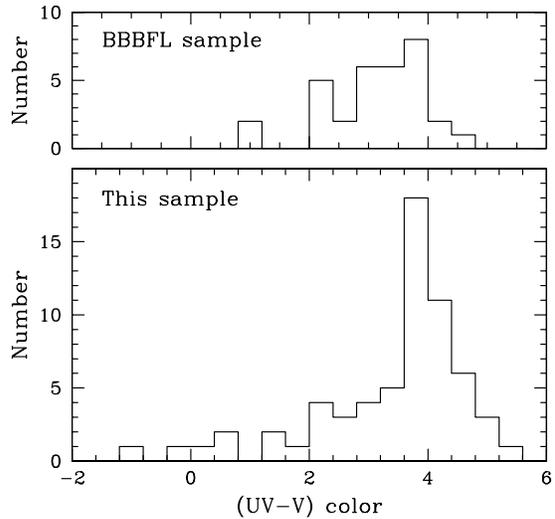}}
  \caption{ Distribution of the UV colors.}  
\end{figure}

\section{ Color-color diagram}

     The scatter of the  $UV-V$ color can be studied as a function of the $B-V$ color
  in the color-color diagram of Figure 2. The objects can be roughly separated in two
  groups. A first group, forming a vertical plume with red $B-V$ colors ($>$ 0.9),
  is consistent with the idea that 
  the population of stars responsible for the UV emission
  and whose changing proportions would explain the scatter in the $UV-V$ color
  are expected to make virtually undetectable contribution 
  at visible wavelengths (O'Connell \cite{oco}).  
  A second group is made of objects that get bluer in $B-V$ and  
  $UV-V$. 
  In addition to star formation, many factors such as the 
 color-magnitude relation and the metallicity
  have already been identified as responsible for
  a scatter (or blueing) in the $B-V$ color of elliptical galaxies. 

   In this presentation we have ignored
  the fact that the $UV-V$ and $B-V$ colors are 
  not completely independent variables. 
  We have verified that the results are similar in 
  the $UV-B$ vs. $B-V$ diagram but we keep using  
  the $UV-V$ color for comparison with the previous work of BBBFL.
  
   In spite of the differences in the observing wavelength and the measurement 
  aperture, the BBBFL data have been plotted in Figure 2 for comparison.
  These data do not 
  populate the diagram in the same specific areas as our data (for instance
  a significant number of objects with $UV-V$ color as blue as 2 but red in $B-V$ 
  are added) but follow the same trends. The $B-V$ colors needed to display the 
  color-color diagram for the BBBFL 
  galaxies have been obtained from the 
  NASA Extragalactic Database. 
  They refer to the integrated light whereas the $UV-V$ colors are for the IUE aperture,
  which may be a problem in some cases.
  These $B-V$ colors have also been found to 
  be on average 0.02 mag bluer than the $B-V$ colors used for our data points and 
  derived from Table 1; this may 
  explain a small horizontal offset between the two datasets in Figure 2.    
  There was no attempt to merge the data into a larger and unique dataset, given 
   the differences described above and the UV-selected origin of our data. 
 
\begin{figure}
  \resizebox{\hsize}{!}{\includegraphics{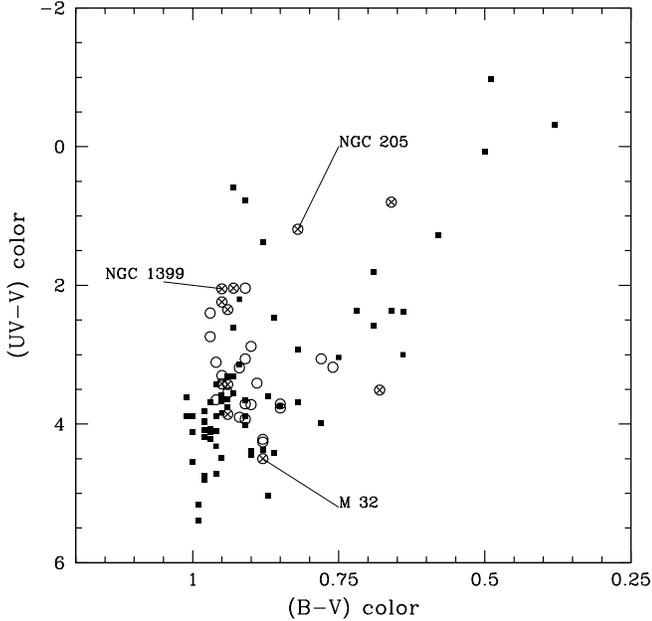}}
  \caption{$UV-V$ vs. $B-V$ color-color diagram for the galaxies of our
   sample (solid square) and the BBBFL sample (open circles). 
   The 11 objects of the BBBFL sample which have been studied 
   in details with HUT or HST and are used as references are marked 
   as diagonal crosses. Great care should be taken when comparing the two samples
   because of differences between the two datasets
   (see text) and the fact that the $B-V$ refers to the integrated light. 
   The specific issue of the $B-V$ of NGC 205 is detailed in subsection 5.4}  
\end{figure}

\section{Analysis}
   
\subsection{Context}

   It is currently thought that HB stars and their progeny are responsible 
  for the far-UV emission in elliptical galaxies. The variety of 
  observed UV spectral energy distributions is explained by changing 
  proportions of stars along the zero-age HB (ZAHB) and 
  the following post-HB evolution tracks.
  Three main classes of post-HB evolution are distinguished,
  each evolving from a different range of effective temperature on the
  ZAHB: from the red end of the HB, the stars evolve as post-AGB stars,
  at hotter temperatures they follow post-early AGB evolution and 
  the bluest follow AGB-Manqu\'e evolution 
  (e.g. Brown et al. \cite{bro2}). The ZAHB location (hence the 
  specific evolutionary track)   
  is driven mainly by 
  the envelope mass: the lower this mass, the hotter is the ZAHB location.
  The envelope mass itself depends critically on the mass 
  losses during the red giant phase. 
  The real populations involve a mixture of these different categories of 
  stars. On one hand, the PAGB-stars fail to  produce enough UV because they
  are UV-bright for a too short period of time; nonetheless they are bright enough
  to be observed. 
  On the other hand, the bluest HB stars and their progeny,  
  if alone, would overproduce UV. The galaxies with stronger UV-upturn
  and bluer UV color have therefore a larger fraction of their 
  populations evolving 
  along these latter paths and a smaller fraction of stars evolving 
  along the PAGB track. 
  Brown et al. (\cite{bro1}) have attempted to place numbers on these 
  fractions. It is generally admitted that 
  the strongest UV-upturn does not require more than about 15 \%
  of the evolving population passing through the hot HB phase. At the opposite,
  PAGB stars alone might account for the weakest UV-upturn. Nevertheless, 
  STIS observations (Brown et al. \cite{bro3}) have shown that the
  hot HB is populated in the case of the weak UV-upturn galaxy M32.
   
  The situation is further complicated by the fact that, even if  
  star formation is ruled out as a general interpretation 
  of the UV emission, it can still be present in a number of instances.
  Features in the optical spectra of some early-type galaxies have been
  interpreted as evidence for 
  a low-level of star formation activity (e.g. Caldwell et al. \cite{cal}); 
  NGC 205 and NGC 5102 are well known
  examples of nearby early-type galaxies with direct evidence for massive 
  star formation (Hodge \cite{hod}, Pritchet \cite{pri}).

\subsection{Comparison with a sample of known objects}
 
  A number of authors 
  have developed spectral population synthesis models, based on the
  different families of evolutionary tracks for low-mass stars
  discussed above
  (e.g. Dorman et al.  
  \cite{dor}, Tantalo et al. \cite{tan}, Yi et al. \cite{yi1}).
  The UV output is found to be extremely sensitive to the parameters used, 
  chief among them the mass loss on the giant branch. Conversely, 
  infering something on the various categories of low-mass stars 
  responsible for the UV flux is difficult, if not impossible, from 
  observations as limited as a broad band measurement.    
  Until the physical processes driving the distribution of 
  stars along the ZAHB and the post-HB evolution are better understood
  our interpretation of UV color will remain limited to analogies
  that can be established  
  with a few objects used as references.

   The BBBFL sample contains a significant number of known objects
   for which a correspondance can be made between the
   $UV-V$ color and the more detailed information available.
   They are NGC 1399 observed with HUT by Ferguson et al. (\cite{fer0}), 
   the 6 early-type galaxies observed with HUT by 
   Brown et al. (\cite{bro0}, \cite{bro1}), 
   M31 and M32 observed with HUT (Ferguson \& Davidsen, \cite{fer}) and 
   the HST (Brown et al. \cite{bro2}, \cite{bro3} and references therein).  
   NGC 205 and NGC 5102 
   have also been studied in details in the UV domain
   (Bertola et al. \cite{ber}, Jones et al. \cite{jon},
   Cappellari et al. \cite{cap}, Deharveng et al. \cite{deh2}).
   All these 11 galaxies are identified 
    in the color-color diagram of Fig. 2.

\subsection{The relative contribution of the 
   different families of low-mass stars in advanced stages of evolution}
    
    For the moment, we concentrate on the spread in $UV-V$ color, irrespectively
    of the scatter in $B-V$, and defer discussion of 
    the bluest objects ($UV-V$ $<$ 1.5)
    close to NGC 205 ($UV-V$ = 1.19) and NGC 5102 ($UV-V$ = 0.8) to  
    the following subsection.
    Most of the objects in Fig. 2 lie in the range of UV colors 2 -- 4.5
   defined by the two extreme objects, M32 ($UV-V$ = 4.5)  and 
   NGC 1399 ($UV-V$ = 2.05) in the small sample of reference ellipticals.
   If the interpretation of NGC 1399 in terms of UV-bright stellar
   content is correct and can be extrapolated to other objects with 
   the same $UV-V$ color, this means that in our relatively large
   sample we have not found 
   galaxies requiring a larger population of hot HB than
   in NGC 1399, i.e. not more than say 15 \% of 
   the evolving population passing through the hot HB. 
   The precision in this number is of course limited by its extreme 
   sensitivity to the envelope mass and the interplay of other 
   parameters (Y, Z, ages).  
   At the opposite,
   the objects as red as 4.5 in $UV-V$ would be explained, if the case 
   of M32 is representative,  
   essentially by PAGB evolution (although population passing
   through the hot HB phase has been resolved 
   in M32 by Brown et al. \cite{bro3}). Of the 7 reddest objects, 
   with $UV-V$ colors
   from 4.5 up to 5.5, 3 are among those in common with the BBBFL sample and 
   reported with a disturbingly redder color than IUE data. 
   These objects are likely to be 
   affected by an integration of the UV flux in a smaller area than in the optical
   band
   and remind us of the uncertainties (including systematics) in the color-color
   diagram. Although uncertain,
   these extreme red UV colors
   would remain compatible with the interpretation by PAGB evolution, 
   given the spread in envelope mass and the role of other parameters
   (Y, Z and ages) 
   as illustrated by  
   the models of Tantalo et al. (\cite{tan}), 
   Yi et al. (\cite{yi1},\cite{yi2}) and references therein.

   An important feature of Figure 2 is the large fraction of galaxies
   with $UV-V$ color in the bin 4 $\pm$ 0.4: they make 29 of the 63 
   galaxies displayed in Figure 2. They outnumber the objects in the 
   bin 2.4 $\pm$ 0.4 by a factor 4,
   or 36 when the difference of 
   volume surveyed (UV-flux limited sample) is roughly accounted for.
   They are not found from the same cluster as the unusually red 
   early-type galaxies reported
   by Marcum et al. (\cite{marc}) in the Perseus cluster.
   If our comparisons above are correct, this dominant number of galaxies 
   with weak UV-upturn would imply that the PAGB stars which are necessarily
   present in elliptical galaxies are the main channel of
   evolution for the low-mass, metal-rich population.
   This large 
   fraction of red objects, implying relatively young ages
   (e.g. Tantalo et al. \cite{tan}, Yi et al. \cite{yi2}), is also consistent
   with the age spreads that are now being reported for early-type 
   galaxies (e.g. Trager et al. \cite{tra}).  
   At the opposite, the scarcity of objects with $UV-V$ in the range 2.5 - 3.2
   suggests a sort of bimodality along the ZAHB or something special
   with the group of six objects (including NGC 1399) reported with 
   a strong UV-upturn with IUE.

   Further interpretation, either on the possibility of a minor contribution of
   stars passing through the hot HB in the reddest galaxies,
   or on the factors controling the distribution at hot temperatures 
   along the ZAHB, would require a larger number of 
   objects and a better photometry than that offered by the present data.

\subsection{The case for recent star formation}
 
   Our sample contains seven objects  
  with UV color bluer than 1.4, comparable to those 
  of NGC 205 ($UV-V$ = 1.19) and NGC 5102 ($UV-V$ = 0.80). 
  By analogy it is tempting to say these seven objects
  also harbour some residual star formation. While this conclusion seems
  reasonable for the bluest of these objects, it is more arbitrary for the 
  others.
  There is indeed no solid argument 
  to draw a line 
  between galaxies with and without residual star formation.   
  It is very possible that an UV color slightly bluer than 2.0
  may be explained by a more extreme than usual 
  mixture of evolved low-mass stars and does not require any additional 
  younger population. Conversely, it is conceivable that a low 
  level of star formation can hide in a few objects 
  with relatively red UV colors. 

  We have examined individually the 7 objects with $UV-V$ $<1.4$ and
  suspected star formation. Hints of star formation are known
  in three of them, CGCG 119053, CGCG 97125 and the bluest VCC 1499 
  (see notes to Table 1). VCC 49 (NGC 4168) has a 
  low-luminosity Seyfert nucleus and CGCG 119030 reminds us that 
  misclassification is also a possibility (see notes to Table 1).
  The two last, VCC 616 and CGCG 119086, have so far nothing special. 
  This analysis reasonably confirms our assertion, based on 
  the comparison with NGC 205 and NGC 5102, that the objects blue in $UV-V$
  (say, $<$ 1.4) may have some residual star formation. It also  
  suggests that the UV light has the potential to sort out   
  new cases of residual star formation.   

  If we now pay more attention to the location of the 
  objects in the color-color diagram of Figure 2,
  three (VCC 616, VCC 49 and CGCG 97125) of the 7 objects with
  $UV-V$ $<1.4$  are puzzling 
  by their relatively red $B-V$ whereas residual star formation is
  expected to move the objects to the blue in both colors. 
  We have no explanations for this situation. 
  The location of NGC 205 also deserves attention: 
  its $B-V$ refers to the 
  integrated light whereas the $UV-V$ from BBBFL refers to the central region; 
  a $B-V$ of the order of 0.65, as expected from the surface photometry
  of Lee (\cite{lee}), would be more appropriate for the comparison. This
  would isolate further the three red ($B-V$) objects discussed above.

     For the 19 galaxies in Table 1 without an $UV-V$ color, 
  the $UV-m_{pg}$ color can be a useful approximation, especially 
  for sorting out blue objects as discussed in this subsection. Of the
  five objects with $UV-m_{pg}$ $<0$ 
  (VCC 1809, VCC 1111, CGCG 98078, VCC 914, VCC 288),
  one again (CGCG 98078) is known to
  have an extremely strong H$\alpha$ emission (see notes to Table 1). 

   It is difficult to translate such signs of on-going star formation into  
  something more quantitative. For one, it is unknown whether the 
  residual star formation takes place in an elliptical with a weak 
  or a strong UV-upturn
  as caused by low-mass evolved stars. 
  Second, the same UV color excess may result from different combinations 
  of burst strength, duration and ages. 
  The situation is illustrated for an instantaneous burst in Figure 3 
  displaying the combinations of strength and age
  that would 
  make the $UV-V$ color of an host elliptical bluer than a given limit
  ($UV-V$ of 1.4 and 1  adopted in Fig. 3). 
  The time evolution of the spectral energy ditribution 
  from the burst has been calculated with 
  STARBURST 99 (Leitherer et al. \cite{lei}). A mass-to-visual light ratio
  of 7 M$_{\sun}$ L$^{-1}_{V, \sun }$ is assumed 
  for the host elliptical (Charlot et al. \cite{cha}, Bressan et al. 
  \cite{bre}),
  allowing us to define the burst strength as a fraction of the mass of the 
  host and to calculate the composite color.  

\begin{figure}
  \resizebox{\hsize}{!}{\includegraphics{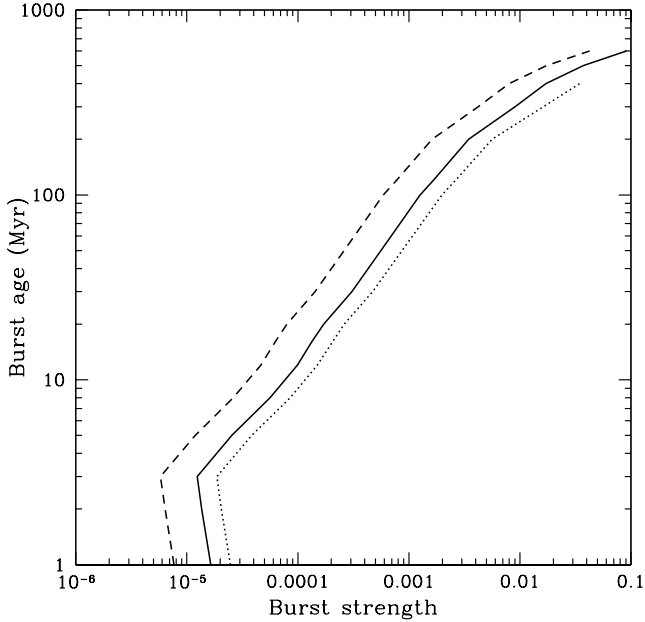}}
  \caption{ The instantaneous bursts able to make an elliptical bluer than
    a given $UV-V$ color have their strengths and ages in the domain below
    and to the right of the curves. Solid line: composite color $<$ 1.4,
    host color = 4; dashed line: same but host color = 2; dotted line:
    composite color $<$ 1, host color = 4. Solar metallicity and 
    a Salpeter IMF (1 -- 100 M$_{\sun}$) have been used in 
     STARBURST 99 calculations.}  
\end{figure}

    If a constant star formation is assumed, it is also possible 
  with STARBURST 99 (Leitherer et al. \cite{lei}) to 
  calculate the rate which would make an host elliptical bluer
  than a given $UV-V$ limit; the result has, however, to be  expressed 
  per unit V-band luminosity and becomes independent of duration for 
  period over 100 Myrs.  
  With the limit of  $UV-V$ = 1.4 as in Fig. 3
  we obtain  $9.2 \times 10^{-12}$ M$_{\sun}$ yr$^{-1}$ L$^{-1}_{V, \sun }$
  and  $4.3 \times 10^{-12}$ M$_{\sun}$ yr$^{-1}$ L$^{-1}_{V, \sun }$
  for an host with a $UV-V$ color of 4 and 2 respectively. For an elliptical
  with $M_V = -21$, this translates into
  rates of 0.2 and 0.09 M$_{\sun}$ yr$^{-1}$ 
  respectively. Such numbers are comparable with 
  the rate of gas shed by stellar evolution in early-type galaxies
  (e.g. Faber \& Gallagher \cite{fab}). As noted by O'Connell
  (\cite{oco}), a complete recycling into new stars is therefore excluded 
  by UV observations as a regular phenomenon in early-type galaxies.
  In contrast,
  the few objects bluer than the limits adopted suggest special events
  triggered by interaction and gas transfer. 
  Constraints on a partial gas recycling would require to appreciate 
  lower level of star formation, which is not yet permitted by the 
  scatter of UV colors and the present understanding of the UV emission
  from HB stars and post-HB progeny.  

\subsection {The scatter in $B-V$ and relationship to other properties}

    An important feature of Figure 2 is the significant number of 
   objects that have relatively blue $B-V$ color (say $<$ 0.85),
   in contrast to the galaxies in the red vertical plume, 
   as exemplified by the 4 reference objects 
   (NGC 1399, NGC 4649, NGC 4552 and 
   NGC 4486) which have red $B-V$ $\sim$ 0.95
   (with $UV-V$ $\sim$ 2.2).

    We have first 
   examined whether a blueing in $B-V$ may be explained by 
   the variety of evolution paths responsible for the scatter of the UV color.
   We have not found any evidence for a blueing and a scatter of the $B-V$   
   color in the current stellar population models reproducing 
   the UV color (Bressan et al. \cite{bre}, Tantalo et al. \cite{tan}).

     Residual star formation 
   is a possible explanation, especially since the correlation
   between the $UV-V$ and $B-V$ extends into 
   the domain of the bluest $UV-V$ galaxies that have been discussed 
   in terms of star formation in the previous subsection. 
   However, because the galaxies blue in $B-V$
   are not all very blue in $UV-V$, 
   it is also reasonable to consider other factors such as 
   the diversity of luminosity and morphological type.
    
    An individual examination of extreme objects is instructive.
   Of the seven objects with $B-V$ $<$ 0.75 (and $UV-V$ $>$ 1.5),
   three are dwarf ellipticals (VCC 608, VCC 781, VCC 951), 
   a category of objects known to have bluer $B-V$ than
   regular ellipticals (e.g. Ferguson, \cite{fer1}). Three are  
   S0 galaxies (CGCG 127032, CGCG 160068, CGCG 160120); 
   among them, CGCG 127032 (NGC 3821) is perhaps misclassified (see note
   to Table 1) and 
   has a significant amount of neutral hydrogen (Eder et al. \cite{ede}),
    CGCG 160068 (NGC 4853) has strong Balmer absorption lines 
   (see note to Table 1) and is the bluest in $UV-V$ of all seven. 
   The last object (of the seven),     
    VCC 1327 (NGC 4486A), should be discarded because a bright star 
   is superimposed and probably contaminates the photometric measurements.

       The potential role of the luminosity (and mass) has also been 
    explored in various color-magnitude diagrams built 
    with our sample. Figure 4 is an example of these diagrams with the H-band 
    luminosity. The main trend of variation shows  
    the blueing (and scatter) of the $B-V$ color 
    with the low-luminosity objects,
    among them the three Virgo dwarf ellipticals
    discussed above. The branch leaving this main trend at log($L_H$) 
    $\sim$ 11 and $B-V$ $\sim$ 0.75 is caused by objects of the Coma 
    cluster with possible misclassification due to the distance and 
    signs of star formation as discussed above. Interestingly, this  
    branch is comparable to that 
    followed by spiral galaxies in a more general study of the  
    photometric and structural properties of 
    galaxies (Scodeggio et al. \cite{sco}). 

        With an interpretation of the scatter in $B-V$ more in terms  
    of luminosity effects than star formation (both are not exclusive), 
    it remains to understand  
    the apparent lack of galaxies blue in
    $B-V$ ($<$ 0.75 as above) and red in $UV-V$ (say, $\sim$ 4).
    This feature contributes to enhance the correlation 
    between  the $UV-V$ and $B-V$ colors reported for a number 
    of objects in the color-color diagram.
    There is probably 
    a selection effect in the sense that a galaxy 
    with $UV-V$ = 4 and a $B-V$ color implying an absolute 
    magnitude $>-17$ would fall below the UV detection 
    limit at the distance of Virgo.

\begin{figure}
  \resizebox{\hsize}{!}{\includegraphics{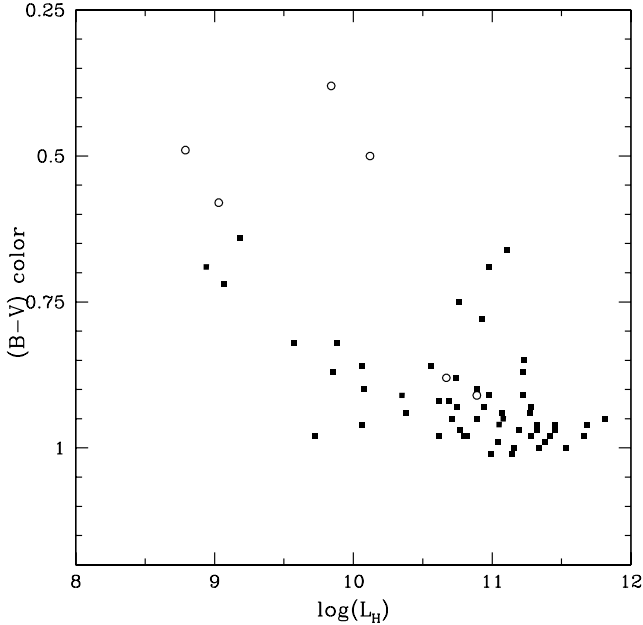}}
  \caption{$B-V$ vs. $H$ color-magnitude diagram for the galaxies of our
   sample (VCC 1327 has been discarded). The objects with $UV-V$ $<$ 1.4
   are plotted as open circles.}  
\end{figure}

\subsection{The Mg$_2$- UV color correlation}

     The correlation found between the UV color and the spectral line 
   index Mg$_2$, with the color being bluer in more metal-rich galaxies
   (Faber \cite{fab1}, BBBFL), has
   played a crucial role in establishing evolved 
   low-mass stars rather than massive stars as the main source of 
   UV emission in early-type systems. In contrast,
   the 42 objects of our sample that have both an $UV-V$ color and 
   a Mg$_2$ index  
   do not show such a correlation (Figure 5).
   There are several possible explanations for this difference.
    
    First, the Mg$_2$ index obtained from the literature 
   refers to the central regions whereas our $UV-V$ color refers to 
   the integrated light. This may account for some of the scatter
   in Fig.5 but not for the lack of correlation, as shown by the
   amplitudes of aperture correction displayed by Golev \& Prugniel
   (\cite{gol}).     

    Second, the Mg$_2$-UV correlation of BBBFL is more apparent 
   (as in Dorman et al. \cite{dor}) after removing
   galaxies with activity or on-going star formation. 
   In such domains of the plot,
   we also have data points that might be removed on the same criteria.      
       
   Third and more important, Fig. 5 shows that our data do not contradict 
   the BBBFL correlation in terms of location in the plot
   but are clumped within a narrower domain than the BBBFL data. 
   This feature, combined with the relatively large dispersion at a given
   Mg$_2$ index, is in agreement with   
   the idea (Dorman et al. \cite{dor}) that the correlation
   may arise from distinct classes of galaxies
   rather than from a continuum of properties. Most of our data
   would belong to an intermediate group with modest dependence of UV colors 
   on Mg$_2$. 
 
\begin{figure}
  \resizebox{\hsize}{!}{\includegraphics{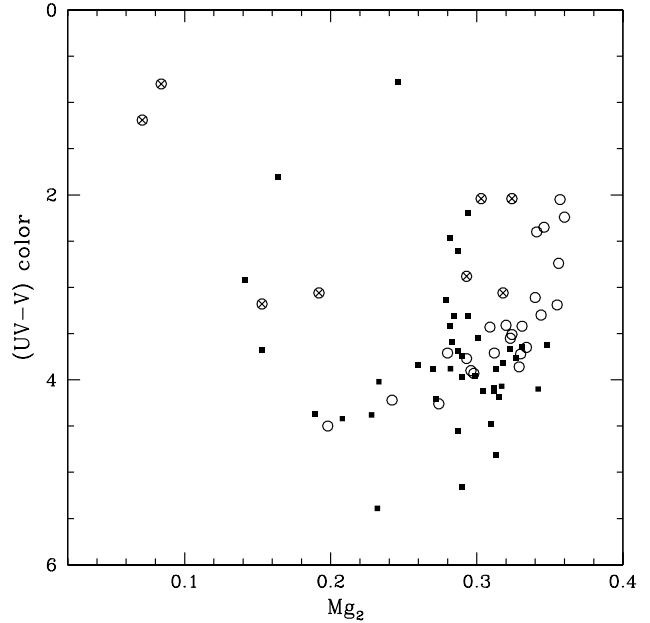}}
  \caption{ The $UV-V$ color as a function of the Mg$_2$ spectral line index.
   The sample of BBBFL is displayed for comparison (circles);  their  
   correlation is enhanced when galaxies with activity or on-going
   star formation (circles with diagonal crosses) are removed.
   The objects redder than 4.5 are among those discussed in section 3 and 5.3
   and may have their colors affected by aperture mismatch.}  
\end{figure}

   At the interpretation level the lack of correlation is not so 
   embarrassing since the dominant factor for 
   the production of UV light is not the metallicity but the
   distribution of envelope masses on the ZAHB, which is itself 
   determined by mass loss on the giant branch.
   The explanation for the correlation
   would be that mass-loss parameter increases with Z 
   (Dorman et al. \cite{dor}, O'Connell \cite{oco}).  
   In contrast, the current correlation of the $B-V$ color with
   metal abundance, driven by opacity effects in stellar atmospheres,
   is equally clear in the two samples with the color getting redder
   as the Mg$_2$ index increases.

\subsection{UV light profile and color gradient}
 
    Four galaxies in our sample have a significant angular extent 
  that allowed us to derive a radial UV profile
  and $UV-B$ color profile, 
  according to procedures described by Gavazzi et al. (\cite{gav6}).
  The angular resolution of $\sim$20\arcsec~ in the UV images prevents 
  any conclusion to be reached within a radius of 20\arcsec, a domain where
  most of the galaxies are known to exhibit a plateau in $UV-B$ color 
  (Ohl et al. \cite{ohl}). Beyond that radius and up to 50\arcsec, 
  the $UV-B$ color of M87 displays a reddening of 0.7 mag comparable 
  to that reported by Ohl et al.
  (\cite{ohl}) and then reaches a plateau up to 
  150\arcsec, beyond the measurements with UIT (Ohl et al. \cite{ohl}).
  In all three other objects (NGC 4374, NGC 4406 and NGC 4429) the  
  angular extent is less and the external data points have low precision;
  NGC 4374 and NGC 4406 at least do not seem to support a reddening 
  as reported by Ohl et al. (\cite{ohl}) in all their objects except M32.

\section{Prospects for the GALEX survey} 

       The future GALEX survey (e.g. Martin et al. \cite{mar}) will 
  considerably increase the number of flux measurements of galaxies
  in the far-UV.
  Adopting the counts of early-type systems of Marzke et al. (\cite{marz}),
  assuming a limiting UV magnitude of 18 and a 
  simplified mixture of ellipticals with two $UV-V$ 
  colors (4 and 2, in proportion 4 to 1 as in the present paper),
  we estimate that GALEX may detect about 4400 early-type
  galaxies in 20,000 square degrees.

   The sheer size of the 
  sample would allow us to explore the role of many parameters 
  in the UV properties 
  themselves. In combination with the uniformity of the UV-flux 
  selection, it should be possible to expand over the 
  two major issues addressed here, the relative proportion of 
  the different categories of evolved stars responsible for the UV emission
  and the relative frequency of residual star formation activity.
  From then and the measurement of the volume surveyed, 
  it should be possible to 
  estimate the volume density of star formation in early-type
  galaxies. Such an evaluation         
  was not possible with our present sample because of the small number 
  of objects, which in addition were mostly cluster galaxies.

   The GALEX survey will also provide  
  spectral information not available in the present sample,
  in the form of two UV bands (and low-resolution spectra for some of the 
  objects). Models based on two UV bands and  
  developed by Dorman et al. (\cite{dor}), Yi et al. (\cite{yi1},
  \cite{yi2}) will help to take advantage of this additional information.

\section{Conclusions}

    We have assembled a sample of 82 early-type galaxies with a
    flux measurement in the far-ultraviolet. In addition to more 
    than doubling the number of objects, 
    this sample has the advantage 
    to be essentially UV-flux selected. 
    The following has emerged from the analysis.

    1) The large scatter of the $UV-V$ color in comparison with the colors
   in the optical is
   confirmed. As shown with a small number of objects 
   studied previously in much detail, the color spread
   between 2 and 5 might be explained by changing proportions 
   of stars along the ZAHB and the following post-HB evolutionary tracks.

    2) The galaxies with red $UV-V$ ($\sim 4$) colors (or weak UV upturn) 
    outnumber those with blue $UV-V$ ($\sim 2$) colors (or strong UV upturn) 
    in our sample.
    If the current interpretation of the UV-upturn 
    can be extended to our 
    sample, the PAGB tracks would be the most common
    evolution path among elliptical galaxies.
    Only a minority of elliptical galaxies would need a fraction of 
    their stars evolving from the blue part of the ZAHB. 
    The GALEX survey should considerably refine this finding, including
    possible differences between the various categories of early-type
    galaxies.          
     
    3) Few blue objects ($UV-V$ $<$ 1.5)  may harbour some residual star 
    formation as shown by the examples of NGC 205 and NGC 5102.
    The implication in terms of the cosmic density of the star 
    formation rate in early-type galaxies should await 
    for a more extended survey like GALEX.     

    4) For a fraction of the objects, the scatter of the $UV-V$ 
    color is accompanied by a 
    scatter in the $B-V$ color. The latter should be caused by the 
    variety of morphological types and luminosities in the sample rather than 
    the evolutionary features explaining the 
    UV emission. 

    5) The correlation between the $UV-V$ color and the Mg$_2$ spectral
    index is not found. This is in line with the idea that the UV flux is 
    not driven by the metallicity but by the mass loss along the giant branch 
    that determines the envelope mass on the horizontal branch.

\begin{acknowledgements}
   We thank B. Milliard, M. Laget and M. Viton 
  for providing new sets of UV fluxes 
  of galaxies in advance of publication, G. Gavazzi for sharing with us 
  a large number of complementary photometric data,  
  and A. Donati for running for us his program on the radial light profiles.
  The referee, R. O'Connell, is thanked for improvements to the text
  and for pointing out interesting features in the data. 
\end{acknowledgements}

\end{document}